\UseRawInputEncoding
\documentclass[pra,twocolumn,showpacs,amssymb,aps,floatfix]{revtex4}
\newcommand{\be}{\begin{equation}}
\newcommand{\ee}{\end{equation}}
\newcommand{\bea}{\begin{eqnarray}}
\newcommand{\eea}{\end{eqnarray}}
\usepackage{graphicx}
\usepackage{amsmath}
\usepackage{color}
\begin{document}

\title{Autonomous stabilization of photonic Laughlin states through
angular momentum potentials}
\author{R. O. Umucal\i lar}
\affiliation{%
Department of Physics, Mimar Sinan Fine Arts University, 34380 Sisli, Istanbul, Turkey
}%
\author{J. Simon}
\affiliation{%
Department of Physics, The James Franck Institute, and the Pritzker School of Molecular Engineering, University of Chicago, Chicago, Illinois 60637, USA
}%
\author{I. Carusotto}%
\affiliation{%
INO-CNR BEC Center and Dipartimento di Fisica, Universit\`a di Trento, 38123 Povo, Italy
}%

\date{\today}
\begin{abstract}
We propose a method to stabilize Laughlin states of a large number of
strongly interacting photons by combining a frequency-selective
incoherent pump with a step-like potential in the angular momentum basis.
Analytical expressions for the preparation efficiency and for the
principal error sources are obtained. Direct extension of the
preparation scheme to states containing single or multiple quasiholes is
discussed.
\end{abstract}


\maketitle

{\it Introduction.---} The impressive recent advances of topological photonics \cite{Ozawa19} are
suggesting photonic systems as a most promising platform to study
fractional quantum Hall liquids \cite{Yoshioka,Cooper08} in a new context that
takes full advantage of the peculiar manipulation and diagnostic tools
offered by optical techniques to investigate the many-body state of the
photon fluid \cite{Caru_RMP13}.
Among the many specific systems that are being investigated to this
purpose, most advanced results have been so far obtained using photonic
lattices in the microwave domain of circuit QED and Rydberg polaritons
in twisted cavities in the visible domain.
On the former platform, chiral motion of strongly interacting photons
under the effect of a synthetic magnetic field has been observed in a
three-site geometry \cite{Rou16} and the autonomous stabilization of a
mesoscopic Mott insulator state was highlighted in a one-dimensional
lattice \cite{Ma19}. On the latter platform, a two-photon Laughlin state was
coherently generated and then studied in its correlation functions
\cite{Clark20}.

In order to be able to explore the peculiar topological properties of
FQH liquids, the most challenging step that remains open concerns the
achievable size of the photon fluid.
Capitalizing on the recent advances, a most exciting perspective is to
merge the autonomous stabilization techniques first proposed in
\cite{Kap14,Leb16} and experimentally developed in the microwave domain in
\cite{Ma19} with the synthetic magnetic field and the strong interactions
developed for Rydberg polaritons in \cite{Clark20} and approach macroscopic system sizes where topological properties become dominant such as quantization of the transverse conductivity and fractionalization of excitation charge and statistics~\cite{Yoshioka}.

In a recent work \cite{Umu17}, two of us proposed the autonomous
stabilization of Laughlin states via a frequency-selective incoherent
pumping scheme suitable for the experimental set-up of \cite{Clark20}. That
study being based on a numerical simulation of the full driven-dissipative
master equation, it was strongly limited in the number of particles
accessible to the calculations and further complications were introduced
by the use of a real-space hard-wall potential to spatially confine the
FQH fluid. In particular, no analytical insight could be offered for the
actual scaling of the preparation efficiency in the interesting regime
of large photon numbers.

In the present work we make a further step in this enterprise by
proposing a new confinement strategy based on a step-like potential in the angular momentum basis. In addition to the simplicity of its experimental realization, this form of confinement potential allows for a full theoretical characterization of the
competing processes due, e.g., to the generation and the subsequent
refilling of quasiholes during photon loss and repumping cycles. The results of numerical simulations for small system sizes can thus be complemented with accurate analytical estimates of the preparation efficiency under a realistic
driven-dissipative protocol. The conclusions of our joint numerical and analytical studies appear to be promising in view of stabilizing macroscopic samples of quantum Hall liquid of light.

{\it Isolated system and shaping its energy levels.---} We describe the fluid  of interacting photons confined in a two-dimensional plane under a uniform and perpendicular synthetic magnetic field with the following second-quantized Hamiltonian written in terms of the bosonic field operator $\Psi(\mathbf{r})$: 
\begin{multline}
\mathcal{H}
=\int\!d^2\mathbf{r}\,\left\{\Psi^\dagger(\mathbf{r})\left[\frac{(-i\hbar\nabla-\mathbf{A}(\mathbf{r}))^2}{2m_{ph}}\,+\hbar\omega_{cav} \right. \right]\Psi(\mathbf{r})\\ \left.+\frac{\hbar g_{nl}}{2}\,\Psi^\dagger(\mathbf{r})\Psi^\dagger(\mathbf{r})
\Psi(\mathbf{r})
\Psi(\mathbf{r})\right\}.
\label{eq:Hamiltonian}
\end{multline}
The single-particle Hamiltonian is given by terms inside the square brackets, where the synthetic magnetic field $\mathbf{B} = B\hat{\mathbf{z}}$ for photons of unit synthetic charge is defined through the magnetic vector potential $\mathbf{A}(\mathbf{r})$, which we take to be in the symmetric-gauge form $\mathbf{A} = B\hat{\mathbf{z}}\times\mathbf{r}/2$. The shift $\omega_{cav}$ to single-particle energies is the natural cavity frequency of the longitudinal mode that we focus on and $m_{ph} = \hbar\omega_{cav}/c^2$ is the effective photon mass that results from confinement along the perpendicular direction $\hat{\mathbf{z}}$. The last term of the Hamiltonian $\mathcal{H}$ given in the second line of Eq. (\ref{eq:Hamiltonian}) describes effective repulsive contact interactions between photons with strength $g_{nl}$, which is determined by the optical nonlinearity of the medium.

The single-particle states of this system are the Landau levels with equally-spaced energies, the separation being $\hbar B/m_{ph} \equiv 2\hbar\omega_{cycl}$. These states are angular momentum eigenstates in our chosen symmetric gauge and the wave function in the lowest Landau level (LLL) with angular momentum $m\hbar$ is given by $\varphi_m(z) = z^{m}e^{-|z|^{2}/2}/\sqrt{\pi m!}$, where $z = (x + iy)/\ell$ is the complex-valued coordinate of the particle and $\ell = \sqrt{\hbar/m_{ph}\omega_{cycl}}$ is the magnetic length. In order to simplify our theoretical description, we work in the LLL approximation, which is valid when the typical interaction energy $v_0 = \hbar g_{nl}/2\pi\ell^2$ corresponding to the lowest Haldane pseudo-potential for the contact interaction is much smaller than the separation between Landau levels: $v_0\ll 2\hbar\omega_{cycl}$. We incorporate the LLL approximation into our calculations by expanding in Eq. (\ref{eq:Hamiltonian}) the field operator $\Psi({\bf r}) = \sum_m\varphi_m(z)a_m$ in the LLL basis, where the operator $a_m$ annihilates a particle with wave function $\varphi_m(z)$. The Hamiltonian becomes
\be \mathcal{H}_{\rm LLL} = \hbar\omega_0\sum_{i}a^{\dagger}_i a_i + \frac{\hbar g_{nl}}{2\ell^2}\sum_{ijkl} V_{ijkl}a^{\dagger}_i a^{\dagger}_j a_k a_l, 
\label{eq:Hamiltonian_LLL}
\ee
where the energy of a photon in the LLL, taking the natural frequency shift into account, is given by $\hbar\omega_0\equiv \hbar(\omega_{cycl}+\omega_{cav})$ and  the overlap integral $V_{ijkl} = \int\varphi^{\ast}_i(z)\varphi^{\ast}_j(z)\varphi_k(z)\varphi_l(z)dzdz^{\ast}$ quantifies the strength of interactions in the different LLL states. 

It is well-known in the FQH context that for a total angular momentum of $L_z = N(N-1)\hbar$ the exact $N$-particle ground state of the Hamiltonian $\mathcal{H}_{\rm LLL}$ is the bosonic $\nu = 1/2$ Laughlin state \cite{Laughlin1983, Paredes2001}
\begin{equation}
\Psi_{\rm FQH}(z_1, \ldots, z_N) \propto \prod_{j<k}(z_j-z_k)^2e^{-\sum_{i = 1}^N|z_i|^2/2},\label{WF_FQH}
\end{equation}
where $z_j$ is the coordinate of the $j$th particle. Together with the Laughlin state, its edge and quasihole excitations with larger total angular momenta form a massively degenerate manifold of states at energy $N\hbar \omega_0$ as these excited states have zero interaction energy. This lowest-energy manifold is separated from other excited states by a gap of the order of $\Delta=v_0$, which is the exact gap for two particles in the LLL approximation. In the following, besides the Laughlin state, we will be interested in quasihole states $\Psi_{(n){\rm QH}}$ containing one quasihole ($n = 1$) and two quasiholes ($n = 2$) centered at the origin, whose wave functions are obtained by multiplying the Laughlin one (\ref{WF_FQH}) by suitable monomials $\prod_{i=1}^N z_i^{n}$.

\begin{figure*}[htbp]
\includegraphics[scale=0.75]{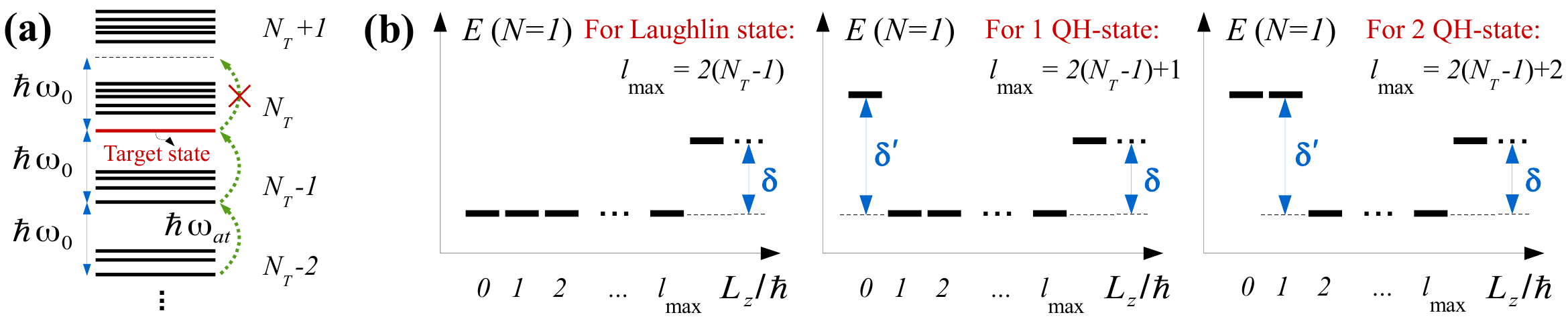}
\caption{(a) Sketch of the general idea of populating a target state with $N_T$ particles through a sequence of intermediate states with equal energy separation $\hbar\omega_0$. Photons with frequency $\hbar\omega_{at} \simeq \hbar\omega_0$ are supplied through a frequency-selective incoherent drive. Excitations above the target state are blocked as they are off-resonant. (b) Sketch of the blocking mechanism via the angular momentum potential with a step-like behavior for different target states. 
\label{Energy level schematics}}
\end{figure*}

The main idea of our proposal to create photonic Laughlin and quasihole states is outlined in Fig. \ref{Energy level schematics}. In order to create these states starting from vacuum, we employ a frequency-dependent incoherent driving scheme. This driving scheme favors upward transitions from an $N$-particle state to an $(N+1)$-particle state compared to the downward ones to an $(N-1)$-particle state as long as the transitions are resonant. As a result, the number of particles in the system keeps increasing until the transition is no longer resonant. As shown in Fig. \ref{Energy level schematics}(a), we take advantage of the equal energy separation $\hbar\omega_0$ between degenerate Laughlin manifolds with successive number of particles to populate a target state with $N_T$ particles by supplying photons with energy $\hbar\omega_{at}$ close enough to $\hbar\omega_0$. What we need to achieve in order to accumulate almost all the population into this target state is to block any further excitation to other states in the same manifold or to the next manifold with $N_T+1$ particles by sufficiently raising the energies of these states so as to make transitions off-resonant. 

As the Laughlin and quasihole states are composed of single-particle states with angular momenta in specific intervals, which are unique to these states, it is possible to raise the energies of the other states in the same degenerate energy manifold and of the ones in the lowest degenerate manifold with one more particle by properly blocking the occupation of single-particle angular momentum modes which lie outside the relevant intervals. For instance, when the target is the $N_T$-particle Laughlin state, which has the smallest total angular momentum in the degenerate manifold, blocking the single-particle states with angular momentum greater than the largest possible angular momentum $l_{\rm max} = 2(N_T-1)$ of a single particle in the Laughlin state [see Fig. \ref{Energy level schematics}(b)] performs the required task as also confirmed numerically. 

In our numerical calculations, this blocking is implemented  by adding to $\mathcal{H}_{\rm LLL}$ an effective angular-momentum potential term which has a simple step-like behaviour:

\begin{multline}
V(N_T,N_{\rm QH}) = \sum_{i}[\delta \, \Theta(i-l_{\rm max}(N_T,N_{\rm QH})-1) \\ 
+\delta^{\prime}\, \Theta(N_{\rm QH}-1-i)]a^{\dagger}_i a_i, 
\label{eq:Angular momentum potential}
\end{multline}
where the Heaviside $\Theta(x)$ function is $0$ ($1$) for $x<0$ ($x \geq 0$), and the number $N_{\rm QH} = 0,1,2$ of quasiholes corresponds to Laughlin, one- and two-quasihole states respectively. The total Hamiltonian is then $H(N_T,N_{\rm QH}) = \mathcal{H}_{\rm LLL}+V(N_T,N_{\rm QH})$. As shown in Fig. \ref{Energy level schematics}(b), the effective potential simply increases by $\delta$ the single-particle energies with angular momenta greater than $l_{\rm max}(N_T,N_{\rm QH}) = 2(N_T-1)+N_{\rm QH}$, that is the largest possible single-particle angular momenta found in the states $\Psi_{\rm FQH}$, $\Psi_{(1){\rm QH}}$, and $\Psi_{(2){\rm QH}}$. Differently from the real-space hard-wall confinement used in~\cite{Umu17} whose effect typically spreads over many angular momentum modes, we consider here a sharp onset of the blocking potential with no disturbance to the Laughlin and the lower states. 
Analogously, as the quasihole $\Psi_{(1){\rm QH}}$ and $\Psi_{(2){\rm QH}}$ states do not contain single-particle modes with $m = 0$ and $m = 0,1$ respectively, they can be selected by raising the energies of the unwanted single-particle modes by $\delta^{\prime}$. 
 
\begin{figure}[htbp]
\includegraphics[scale=0.68]{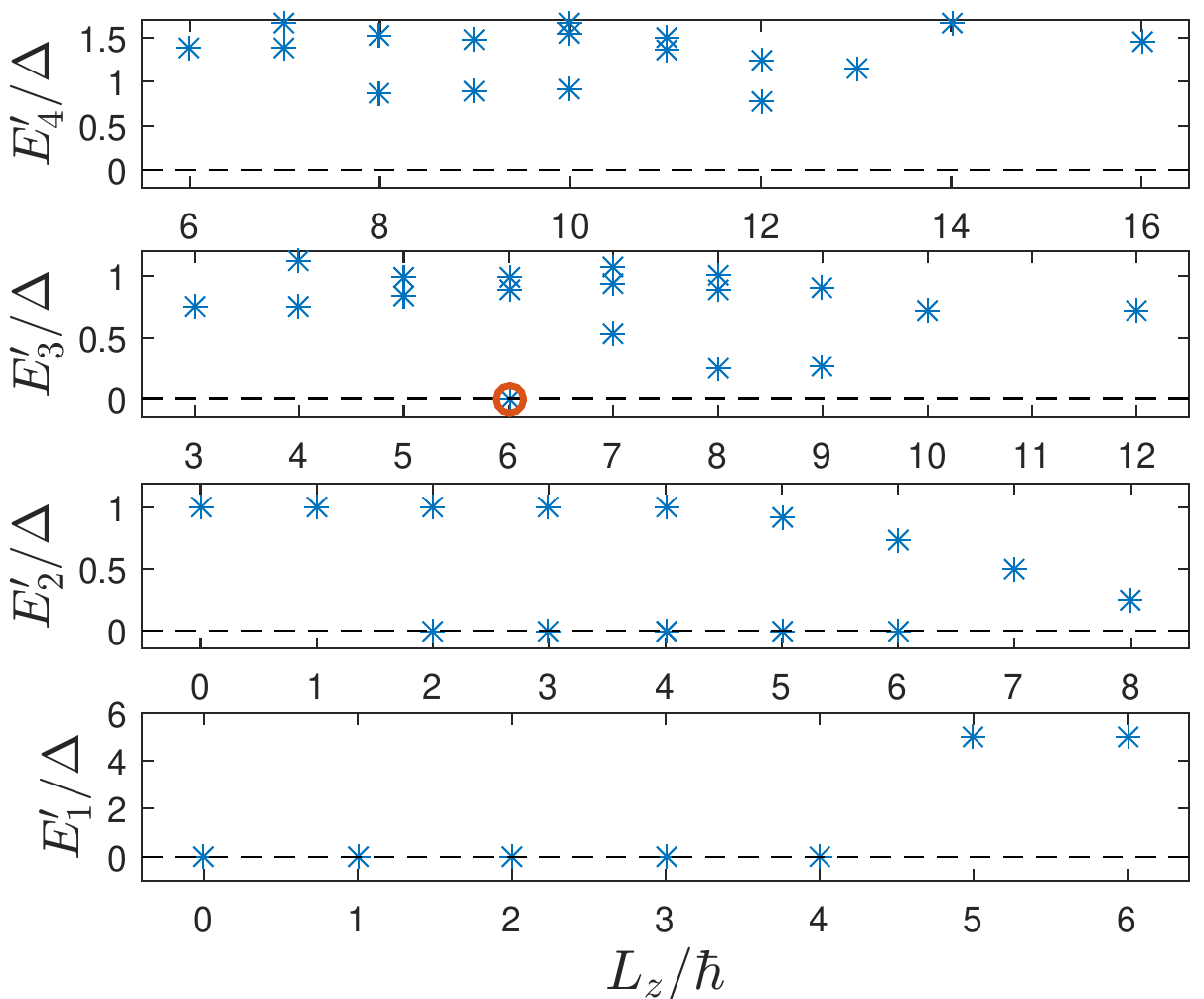}
\caption{$N$-particle energy levels $E^{\prime}_N \equiv E_N-N\hbar\omega_0$ versus total angular momenta $L_z$ when the target state (encircled by a red circle) is the $N_T = 3$-particle Laughlin state, for which we take $l_{\rm max} = 2(N_T-1)$ and $N_{\rm QH} = 0$ in Eq. (\ref{eq:Angular momentum potential}). The angular momentum potential strength is $\delta=5\Delta$. Dashed lines show the non-interacting energy levels in the absence of the potential.
\label{Target N = 3 Laughlin spectra}}
\end{figure}

As a concrete example of the above idea of singling out the target state as the topmost available state in the rung, we show in Fig. \ref{Target N = 3 Laughlin spectra} the many-particle energy levels for the case where we aim at the $N_T = 3$-particle Laughlin state. It is seen that the energy of the two-particle manifold is not shifted up to and including the two-particle two-quasihole state with total angular momentum $L_z/\hbar = 6$, which serves as a leverage for the resonant excitation of the three-particle Laughlin state. For the chosen strength $\delta=5\Delta$ of the angular momentum potential, the target state is seen to be separated from the next states with larger total angular momenta by a gap $\sim 0.25\Delta$ and all the lowest-energy four-particle states are lifted by $\sim 0.80\Delta$ with respect to the value in the absence of the effective potential.

{\it Losses and Incoherent Pumping.---} Here, we briefly discuss how we incorporate the inevitable photon losses and the specific photon replenishing mechanism to our model (cf. the Supplemental Material for details). As discussed in detail elsewhere \cite{Leb16}, the main merit of the specific incoherent driving protocol under consideration is its frequency selectivity, which can be achieved by placing many population-inverted two-level emitters of transition frequency $\omega_{at}$ inside the cavity to obtain a Lorentzian emission spectrum centred around this frequency. If the pumping rate $\Gamma_p$ for the emitters is much larger than the Rabi frequency  of the cavity field-emitter coupling and the spontaneous decay rate, the emitters will most of the time be found in their excited state allowing one to write a master equation only for the photonic density matrix $\rho$ after tracing out the emitter degrees of freedom. This master equation is composed of three parts
\bea
\frac{\partial \rho}{\partial t}\ = -\frac{i}{\hbar}[H(N_T,N_{\rm QH}),\rho]+\mathcal{L}_l+\mathcal{L}_e,
\label{Master}
\eea
where the commutator corresponds to the unitary evolution of the photonic Hamiltonian $H(N_T,N_{\rm QH})$ and the photon losses with rate $\Gamma_l$ are described by a standard Lindblad superoperator $\mathcal{L}_l$. The frequency-selective emission processes are accounted for in terms of a generalized superoperator $\mathcal{L}_e$, which includes appropriately modified field operators. When the emitter transition frequency $\omega_{at}$ matches the frequency difference between two many-particle states with successive number of particles, the emission rate attains a maximum value $\Gamma_e$.
Otherwise, the emission rate is suppressed following a Lorentzian lineshape of linewidth $\Gamma_p$. This frequency-selectivity of the emission process is very well suited for our purpose of populating a specific $N$-particle Laughlin or quasihole state as there are intermediate states starting from the vacuum with successive number of particles whose energies are equally separated by $\hbar\omega_0$, while transitions to undesired states are off-resonant and therefore these states cannot be reached.

{\it Results.---} The master equation (\ref{Master}) can be numerically solved for the steady-state density matrix $\rho_{SS}$.
When the target is an $N_T$-particle Laughlin state, all the (degenerate) lowest-energy non-interacting states with the same number of particles $N<N_T$ turn out to be equally populated provided that $\Gamma_p$ is sufficiently small and $\omega_{at} = \omega_0$. This was anticipated in~\cite{Kap14,Leb16} and is the starting point for analytical considerations in this limit. 

Using the detailed balance condition 
for the populations of states 
in the degenerate manifolds $P_{N+1}^{(0)}\Gamma_l = P_{N}^{(0)}\Gamma_e$ with $N = 0,1,\ldots,N_T-1$, and assuming that only these lowest-energy non-interacting states are occupied appreciably, the target population can be found to be $P_{N_T} \equiv P_{N_T}^{(0)}= 1/[ 1 + \sum_{q = 1}^{N_T}{ d(N_T,q)G^q }]$, where $d(N_T,q)$ is the multiplicity of degenerate states with $N_T-q$ particles and $G = \Gamma_l/\Gamma_e$. 

\begin{figure}[htbp]
\includegraphics[scale=0.56]{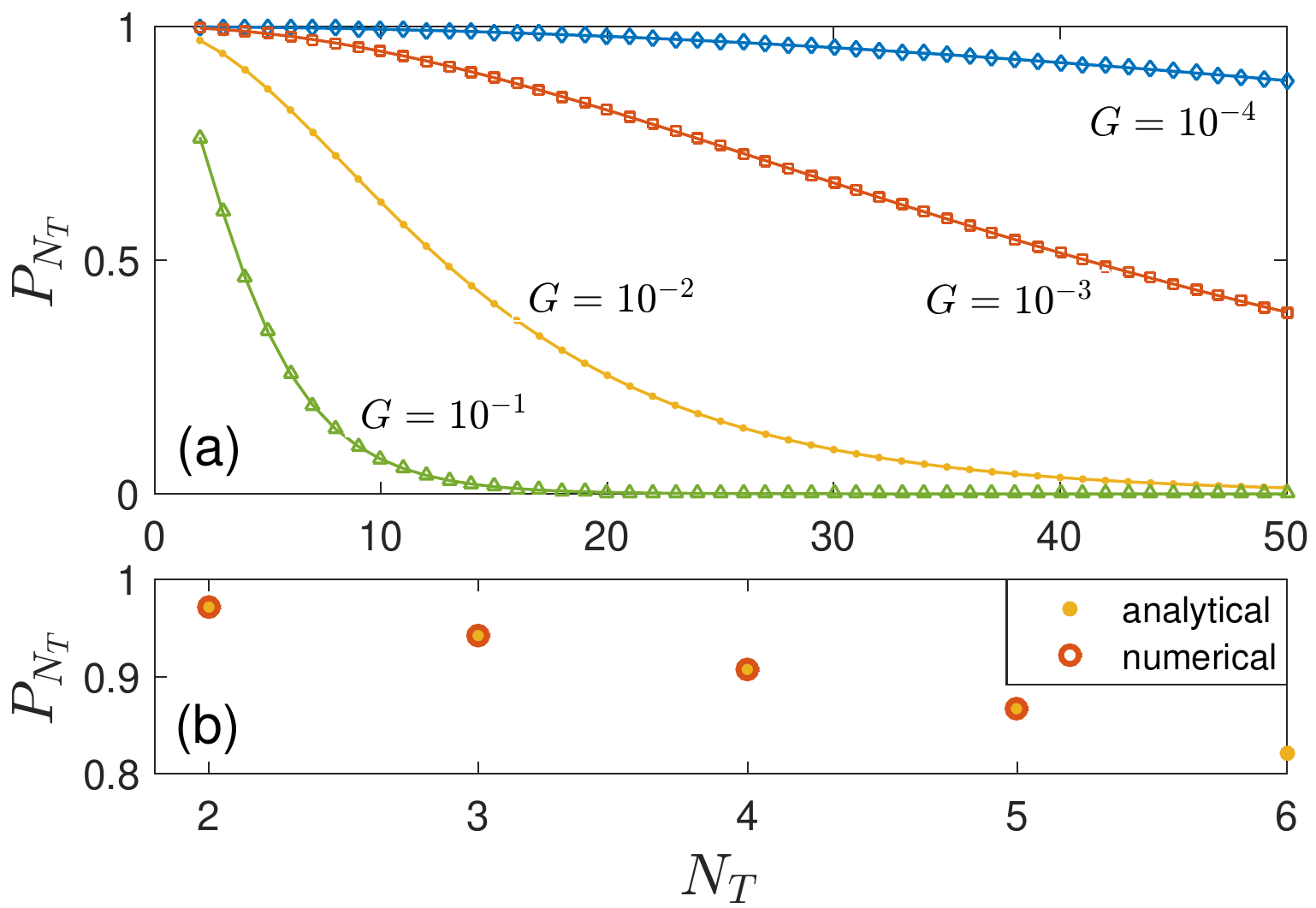}
\caption{(a) Analytical Laughlin state populations (\ref{Analytical Population}); lines are guide for the eye. (b) Comparison of analytical and numerical Laughlin state populations for $N_T = 2,3,4,5$ with $\hbar\Gamma_p/\Delta = 5\times10^{-4}$ and $\Gamma_l/\Gamma_e = 10^{-2}$.
\label{Laughlin populations}}
\end{figure}

The multiplicity $d(N_T,q)$ obtained from the diagonalization of the isolated system Hamiltonian can be accounted for by using a heuristic generalized Pauli principle \cite{Bernevig2008} as follows. This principle asserts for the case of $\nu = 1/2$ that the $N_T$-particle Laughlin state is a superposition of certain states which can be derived from the root state $|\rm R\rangle = |1 0 1 0 1 0 \ldots \rangle$ in the LLL occupation-number representation with a total of $N_T$ occupied orbitals, through an operation called squeezing. Similarly a one- (two-) quasihole state can be created starting from a state containing one (two) extra empty orbital(s) inserted anywhere in $|\rm R\rangle$. Since the loss of a particle from the target state can be thought of being equivalent to creating two quasiholes, the number of possible states reachable from the target state through the loss of $q$ particles can be calculated by counting the number of unique ways of reordering $N_T-q$ times the $(10)$ sequence and $2q$ empty orbitals in a string, yielding the multiplicity $d(N_T,q) = \big(\begin{smallmatrix} N_T+q\\ 2q \end{smallmatrix}\big)$ as a binomial coefficient. Quite remarkably, the sum in the expression for $P_{N_T}$ is explicitly calculable yielding the final result
\be
P_{N_T}= \dfrac{\sqrt{4+G}}{2{\rm cosh}[(1 + 2N_T){\rm asinh}(\sqrt{G}/2)]}.
\label{Analytical Population}
\ee
which is validated in Fig. \ref{Laughlin populations}(b) by comparing its prediction to the numerical results that can be obtained from the master equation (\ref{Master}) for small values of $N_T$. An excellent agreement is found. 

The analytical prediction (\ref{Analytical Population}) is  plotted in Fig. \ref{Laughlin populations}(a) as a function of the particle number $N_T$ for different loss to emission ratios $G$. For a relatively large $G$ ($\sim0.1$) the decay of $P_{N_T}$ is seen to be fast. Indeed in this case the large $N_T$ behaviour is exponential with $\sqrt{4+G}\exp(-2AN_T)$, where $A = {\rm asinh}(\sqrt{G}/2)$. However, for small enough $G$, there is a wide range of photon numbers where $P_{N_T}$ displays a slower decay as $2/(2+GN_T^2)$ as long as $N_T\sqrt{G}$ remains small. Such a slower scaling is greatly conducive to a possible experimental realization of a macroscopically occupied Laughlin state with a large number of particles.

\begin{figure}[htbp]
\includegraphics[scale=0.54]{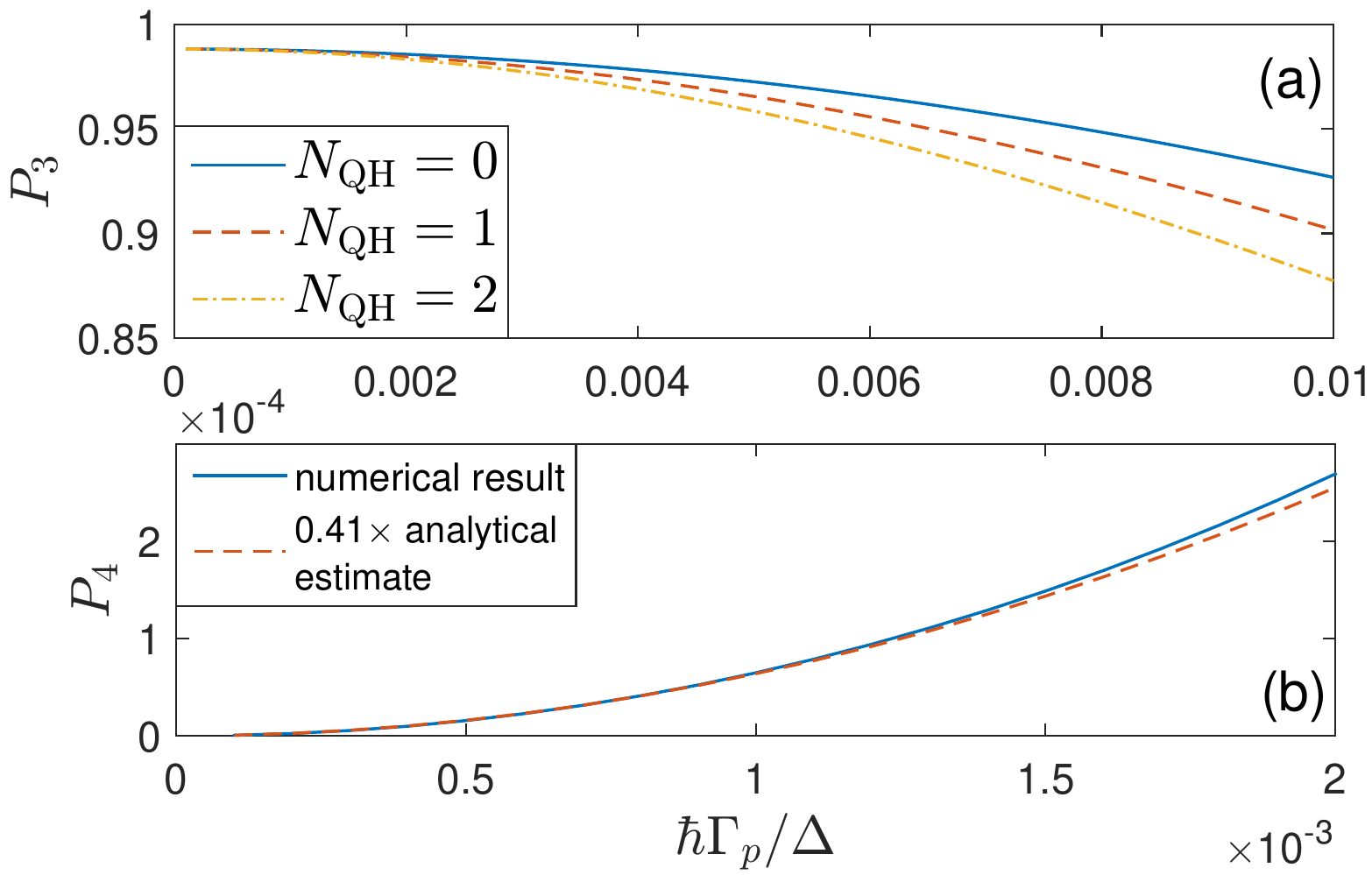}
\caption{(a) Laughlin state ($N_{\rm QH} = 0$), one-quasihole state ($N_{\rm QH} = 1$) and two-quasihole state ($N_{\rm QH} = 2$) populations for $N_T = 3$ as a function of $\hbar\Gamma_p/\Delta$. (b) Comparison between the numerical result for the average population in the $N_T+1=4$ particle states with $\Delta^{\prime} = 0.89\Delta$ as extracted from Fig.~\ref{Target N = 3 Laughlin spectra} for the $N_{\rm QH} = 0$ case and the analytical trend with $\propto \Gamma_p^2$. For both panels, $\hbar\Gamma_e/\Delta = 5\times10^{-5}$, $\Gamma_l/\Gamma_e = 2\times10^{-3}$ and the confinement has $\delta=\delta'=5\Delta$.
\label{Populations for N = 3}}
\end{figure}

Besides the loss to lower degenerate manifolds which is the main reason of the fidelity decrease that is visible in Fig. \ref{Laughlin populations}, another source of decrease is the loss to states lying outside of the degenerate manifolds as the pump linewidth $\Gamma_p$ is increased. Based on a semi-quantitative detailed balance condition between the manifold of non-interacting states and the lowest interacting states (corresponding, e.g., to an extra quasi-particle), we can expect a behaviour $P_{N_T+1} \propto P_{N_T} \Gamma_e/\{\Gamma_l[1+(2\Delta^{\prime}/\hbar\Gamma_p)^2]\}\approx P_{N_T} (\Gamma_e/\Gamma_l)(\hbar\Gamma_p/2\Delta^{\prime})^2$. Here, $\Delta^{\prime}$ is the energy shift of the interacting states, as estimated from the interaction energy of the lowest $N_T+1$-particle states. For the numerically accessible $N_T=3$ value, this behaviour is well confirmed by the full numerics, as shown in Fig. \ref{Populations for N = 3}(b).

{\it Quasihole states.---} We now demonstrate that if a Laughlin state can be created with high fidelity, creating one- and two-quasihole states is almost as effective. In Fig. \ref{Populations for N = 3}(a), we show the three-particle populations in the presence of an additional potential of the form (\ref{eq:Angular momentum potential}) pinning the quasiholes at the center of the FQH fluid. When the pump linewidth $\Gamma_p$ is sufficiently small, the population of the target state turns out to be very similar independently of the number $N_{QH}=0,1,2$ of quasiholes. This basically originates from a very similar structure of the low-energy levels (including their multiplicities) that emerge when these different states are targeted (cf. the Supplemental Material). As $\Gamma_p$ increases, however, differences in the energy-level structure become more pronounced and the smaller energy gap of quasihole states leads to somewhat lower populations for these states compared to the Laughlin one.  

{\it Experimental remarks.---} As a final point, we comment on the actual experimental realization of the step-like potential in the angular momentum basis. In~\cite{Maca17}, it was pointed out that a hard-wall real-space potential had to be very strong and be located very far away from the cloud to provide a step-like dependence guaranteeing the effective upward travel through all the $N<N_T$ states and blocking of the undesired states. While this strategy may be not viable in concrete experimental realizations, an alternative way of designing arbitrary angular-momentum-dependent frequency shifts (\ref{eq:Angular momentum potential}) is based on coupling our main cavity to additional cavities with the same cylindrical symmetry, whose resonant mode pattern can be widely tailored via the length and/or the twist and/or the time-modulation of the cavity, as discussed in the Supplemental Material. This provides a way to restrict the quasi-resonant coupling of the two cavities to specific angular momentum values only, so to engineer the angular-momentum-dependence of the resulting frequency shift of the main cavity's modes.
In this way, the desired step-like potential can be constructed by suitably tailoring a sufficient number of additional cavity modes. Interestingly, efficient stabilization of the Laughlin states only requires blocking the single-particle states around $l_{\rm max}$ and this will automatically prevent  population transfer to higher states as well, which is a further experimental advantage. 

{\it Conclusion.---} We have reported a theoretical study of driven-dissipative fractional quantum Hall fluids of light confined by a step-like potential in the angular momentum basis.
This potential allows for an analytical treatment of the steady-state
solution of the master equation describing the interplay of a frequency-selective incoherent pump with the photon losses. This provides analytical insight on the efficiency
of the autonomous preparation scheme and on the
main sources of error. Our analysis leads to promising conclusions in view of the experimental realization of quantum Hall fluids containing a macroscopic number of particles.

With an appropriate design of the potential, our scheme can be directly extended
to the preparation of single or multiple quasihole states. In this way, 
it can be combined with recent proposals \cite{Umu18,Maca19} for extracting
the fractional statistics from the density profile of the fluid. The fact that edge excitations are gapped by the angular momentum potential and are thus immune to spurious excitations facilitates the assessment of the exclusion statistics via the spectroscopic method proposed in \cite{Cooper+Simon15,Umu17}.
Future work will address time-dependent problems related to the kinetics
of preparation of the Laughlin state starting from vacuum.

R.~O.~U. and I.~C. acknowledges financial support through the T\"UB\.{I}TAK-CNR International
Bilateral Cooperation Program 2504 (project no.~119N192). R.~O.~U  acknowledges financial support through the T\"UBA-GEB\.{I}P Award of the Turkish Academy of Sciences. J.~S. acknowledges support from AFOSR grant FA9550-18-1-0317, and AFOSR MURI FA9550-19-1-0399. I.~C. acknowledges financial support from the H2020-FETFLAG-2018-2020
project "PhoQuS" (n.820392) and from the Provincia Autonoma di Trento.

\end{document}


\title{Supplemental Material for ``Autonomous stabilization of photonic Laughlin states through
angular momentum potentials"}
\maketitle
\section{Energy spectra for target quasihole states}

We carried out a sparse diagonalization of $H(N_T,N_{\rm QH})$ for each particle-number sector to obtain a limited number of eigenstates with lowest eigenenergies in order to maximize the number of particles we can work with. Specifically, we chose this number as the smaller of the total number of eigenstates in a given particle-number sector and 100, a number which we observed in our simulations to be sufficient to accurately obtain populations. This approximation is justified by the fact that our driving protocol is a frequency-selective one which appreciably populates only the low-energy states of each particle-number sector. By choosing the angular momentum cutoff for the single-particle basis states as $m = 2N_T+N_{\rm QH}$, we aimed at including the $(N_T+1)$-particle state containing $N_{\rm QH}$ quasiholes into our simulation as well, when the target state is the $N_T$-particle state containing $N_{\rm QH}$ quasiholes.

\begin{figure}[htbp]
\includegraphics[scale=0.68]{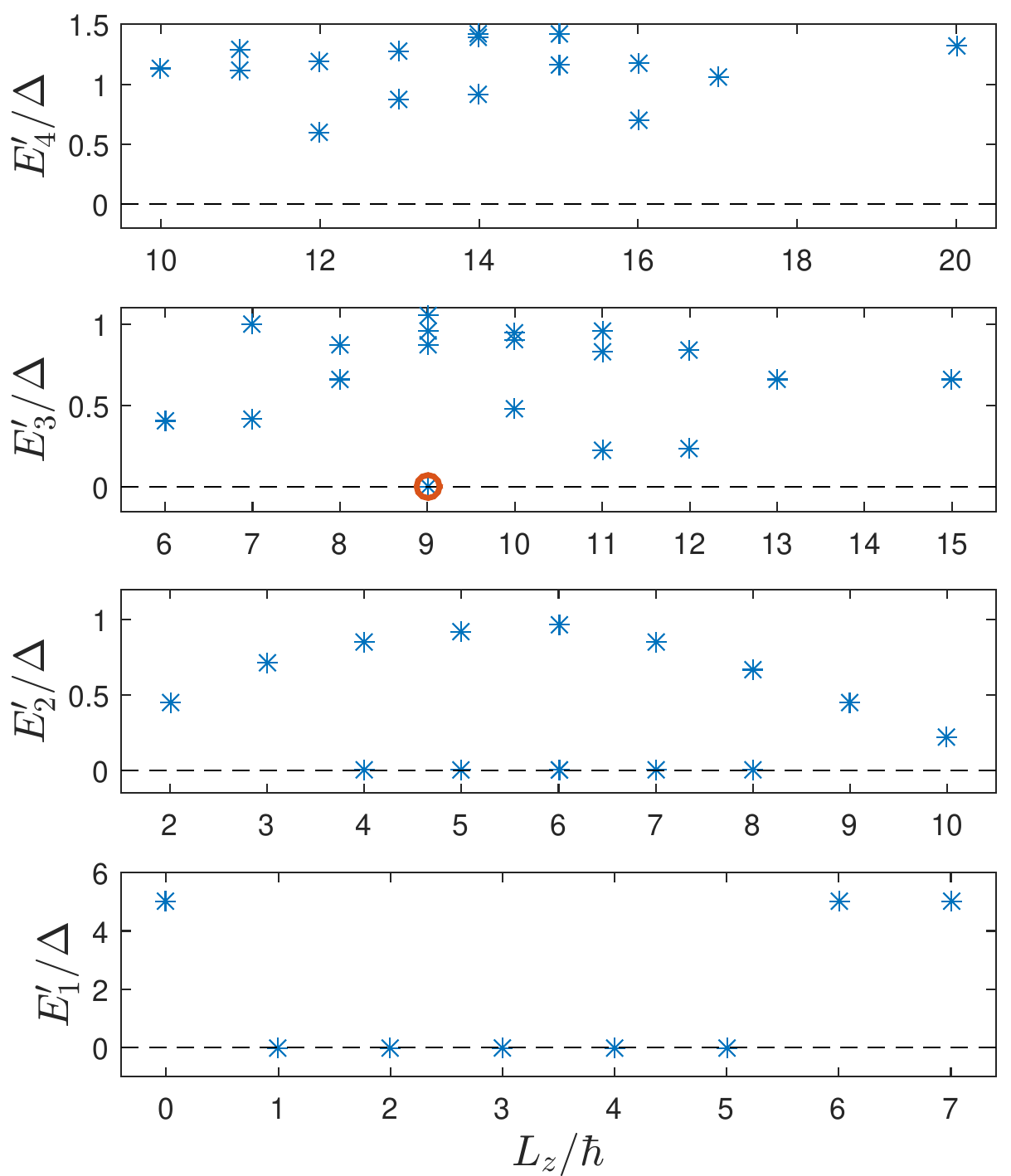}
\caption{$N$-particle energy levels $E^{\prime}_N \equiv E_N-N\hbar\omega_0$ (in units of the typical interaction energy $\Delta$) versus total angular momenta $L_z$ (in units of $\hbar$), when the target state (encircled by a red circle) is the $N_T = 3$-particle one-quasihole state, for which we take $l_{\rm max} = 2(N_T-1)+1$ and $N_{\rm QH} = 1$ in Eq. (6) of the main text. Dashed line shows the unperturbed energy level for the degenerate manifold of the Laughlin state and its edge and quasihole excitations before the application of the angular momentum potential which is given by $E^{\prime}_1$, with strength $\delta = \delta^{\prime} = 5\Delta$.
\label{Target N = 3, QH = 1 spectra}}
\end{figure}

Here we show in Figs. \ref{Target N = 3, QH = 1 spectra} and \ref{Target N = 3, QH = 2 spectra} the energy spectra obtained through such a restricted diagonalization, when the target states are the $N_T = 3$-particle one-quasihole state and the $N_T = 3$-particle two-quasihole state, respectively. Note that the two spectra are very similar almost up to a shift in $L_z/\hbar$ by $N$ in each particle-number sector.

\begin{figure}[htbp]
\includegraphics[scale=0.68]{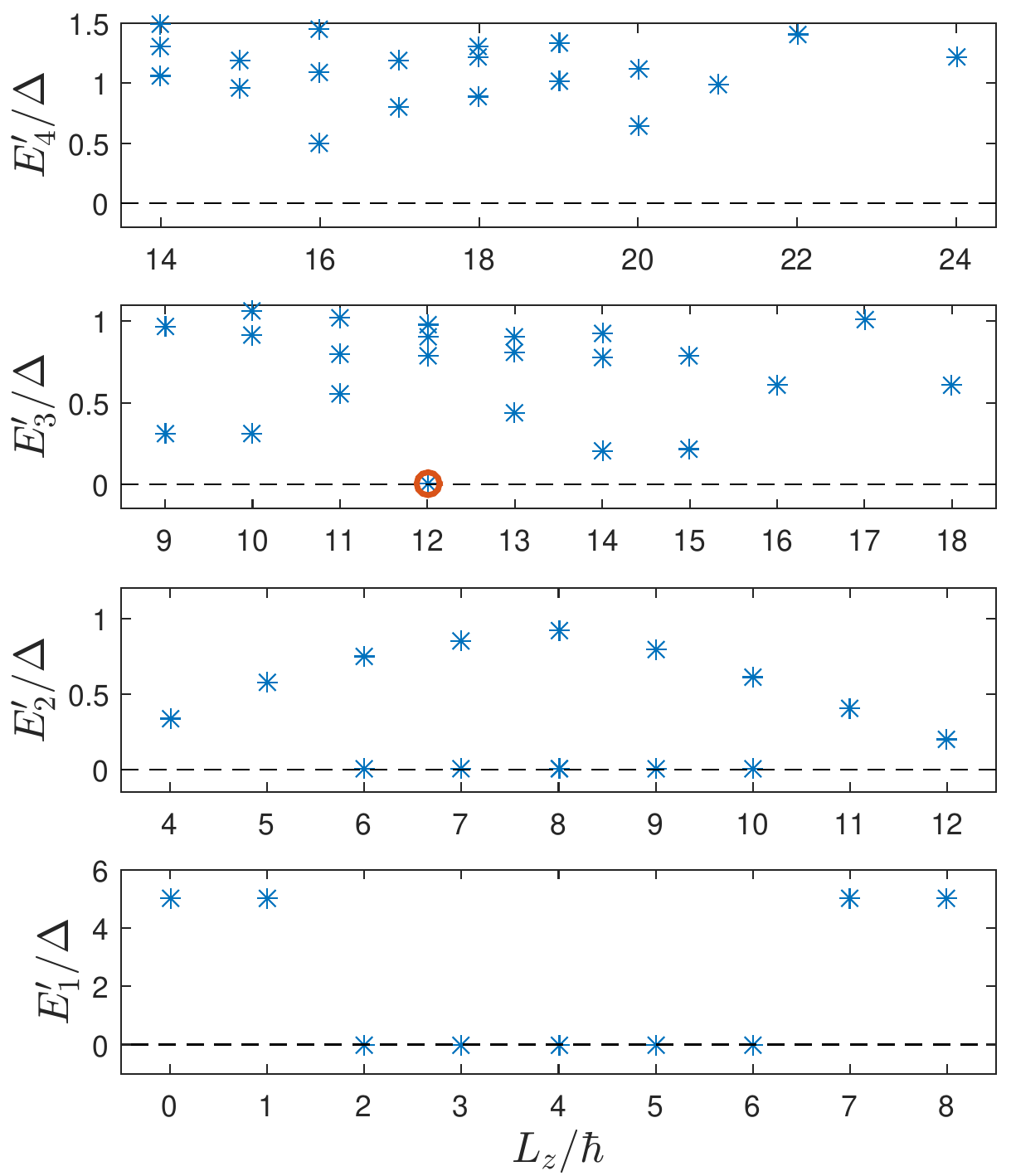}
\caption{$N$-particle energy levels $E^{\prime}_N \equiv E_N-N\hbar\omega_0$ (in units of the typical interaction energy $\Delta$) versus total angular momenta $L_z$ (in units of $\hbar$), when the target state (encircled by a red circle) is the $N_T = 3$-particle two-quasihole state, for which we take $l_{\rm max} = 2(N_T-1)+2$ and $N_{\rm QH} = 2$ in Eq. (6) of the main text. Dashed line shows the unperturbed energy level for the degenerate manifold of the Laughlin state and its edge and quasihole excitations before the application of the angular momentum potential given by Eq.(4) of the main text, with strength $\delta = \delta^{\prime} = 5\Delta$.
\label{Target N = 3, QH = 2 spectra}}
\end{figure}

\section{Losses and incoherent pumping}

The master equation we use is composed of three parts
\bea
\frac{\partial \rho}{\partial t}\ = -\frac{i}{\hbar}[H(N_T,N_{\rm QH}),\rho]+\mathcal{L}_l+\mathcal{L}_e,
\label{Master}
\eea
where the commutator corresponds to the unitary evolution of the photonic Hamiltonian $H(N_T,N_{\rm QH})$ and the photon losses with rate $\Gamma_l$ are described by the Lindblad superoperator
\begin{multline}
\mathcal{L}_l= \frac{\Gamma_l}{2}\int\!d^2\mathbf{r}\,\left[2\Psi(\mathbf{r}) \rho \Psi^\dagger(\mathbf{r})\right.\\
\left.-\Psi^\dagger(\mathbf{r})\Psi(\mathbf{r})\rho - \rho\Psi^\dagger(\mathbf{r})\Psi(\mathbf{r})\right].
\label{Loss}
\end{multline}

The frequency-selective emission processes are accounted for in terms of the generalized superoperator
\begin{multline}
\mathcal{L}_e= \frac{g_e}{2}\int\!d^2\mathbf{r}\,n_{at}\,\left[\tilde{\Psi}^\dagger(\mathbf{r}) \rho \Psi(\mathbf{r})+\Psi^\dagger(\mathbf{r}) \rho \tilde{\Psi}(\mathbf{r})\right.\\
\left.-\Psi(\mathbf{r})\tilde{\Psi}^\dagger(\mathbf{r})\rho - \rho\tilde{\Psi}(\mathbf{r})\Psi^\dagger(\mathbf{r})\right],
\label{Emission}
\end{multline}
where we assume the density $n_{at}$ of the atomic emitters to be constant in the region of interest and $g_e = 4\,\omega_{cav}\,|d_{eg}|^2/(\hbar\Gamma_p L_{\perp})$ is the coupling of each emitter to radiation in terms of the dipole matrix element $d_{eg}$, the cavity length $L_{\perp}$ and the pumping rate $\Gamma_p$. 

The modified field operator $\tilde{\Psi}(\mathbf{r})$ is defined as
\begin{equation}
\tilde{\Psi}(\mathbf{r}) = \frac{\Gamma_p}{2}\int_0^{\infty}\!d\tau\, e^{(-i\omega_{at}-\Gamma_p/2)\tau}\Psi(\mathbf{r},-\tau),
\label{Modified field}
\end{equation}
where we have assumed that the loss rate $\Gamma_l$ and the effective emission rate $\Gamma_e \equiv g_e n_{at}$ are much smaller than the repumping rate $\Gamma_p$. The condition $\Gamma_e\ll \Gamma_p$ directly follows from the initial hypothesis $\Omega_R\ll \Gamma_p$ required to exclude collective Rabi oscillations in favor of an irreversible emission process. These approximations lead to
\begin{equation}
\Psi(\mathbf{r},\tau) \simeq e^{i\mathcal{H}\tau/\hbar}\Psi(\mathbf{r})e^{-i\mathcal{H}\tau/\hbar},
\label{Time dependent field}
\end{equation}
which only depends on the Hamiltonian evolution.

The Hermitian conjugate of Eq.~(\ref{Modified field}) together with Eq.~(\ref{Time dependent field}) yield the matrix elements of the modified field operator in the eigenstate basis $|N\rangle$ of $H(N_T,N_{\rm QH})$ as
\begin{multline}
\langle N+1|\tilde{\Psi}^\dagger(\mathbf{r})|N\rangle = \frac{\Gamma_p/2}{-i(\omega_{at}-\omega_{N+1,N})+\Gamma_p/2}\\
\times\langle N+1|\Psi^\dagger(\mathbf{r})|N\rangle,
\label{Modified original relation}
\end{multline}
where $\omega_{N+1,N} \equiv (E_{N+1}-E_N)/\hbar$ is the frequency difference between two many-particle eigenstates with $N$ and $N+1$ number of particles. The frequency-selectivity of the emission can be understood by inspecting the effective emission rate for a $|N\rangle \rightarrow |N+1\rangle$ transition, which is described by the real part of the factor connecting the matrix elements in Eq.~(\ref{Modified original relation}). When the emitter transition frequency $\omega_{at}$ matches the frequency difference $\omega_{N+1,N}$, the emission rate attains its maximum value $\Gamma_e=g_e n_{at}$. Otherwise, the emission rate is suppressed following a Lorentzian lineshape of linewidth $\Gamma_p$.

In the numerical solution of the master equation we made use of the block-diagonal form of the steady-state density matrix $\rho_{SS}$ in the $N$ and $L_z$ sectors to reduce the computational effort.

\section{Loss channels}

\begin{figure}[htbp]
\includegraphics[scale=0.57]{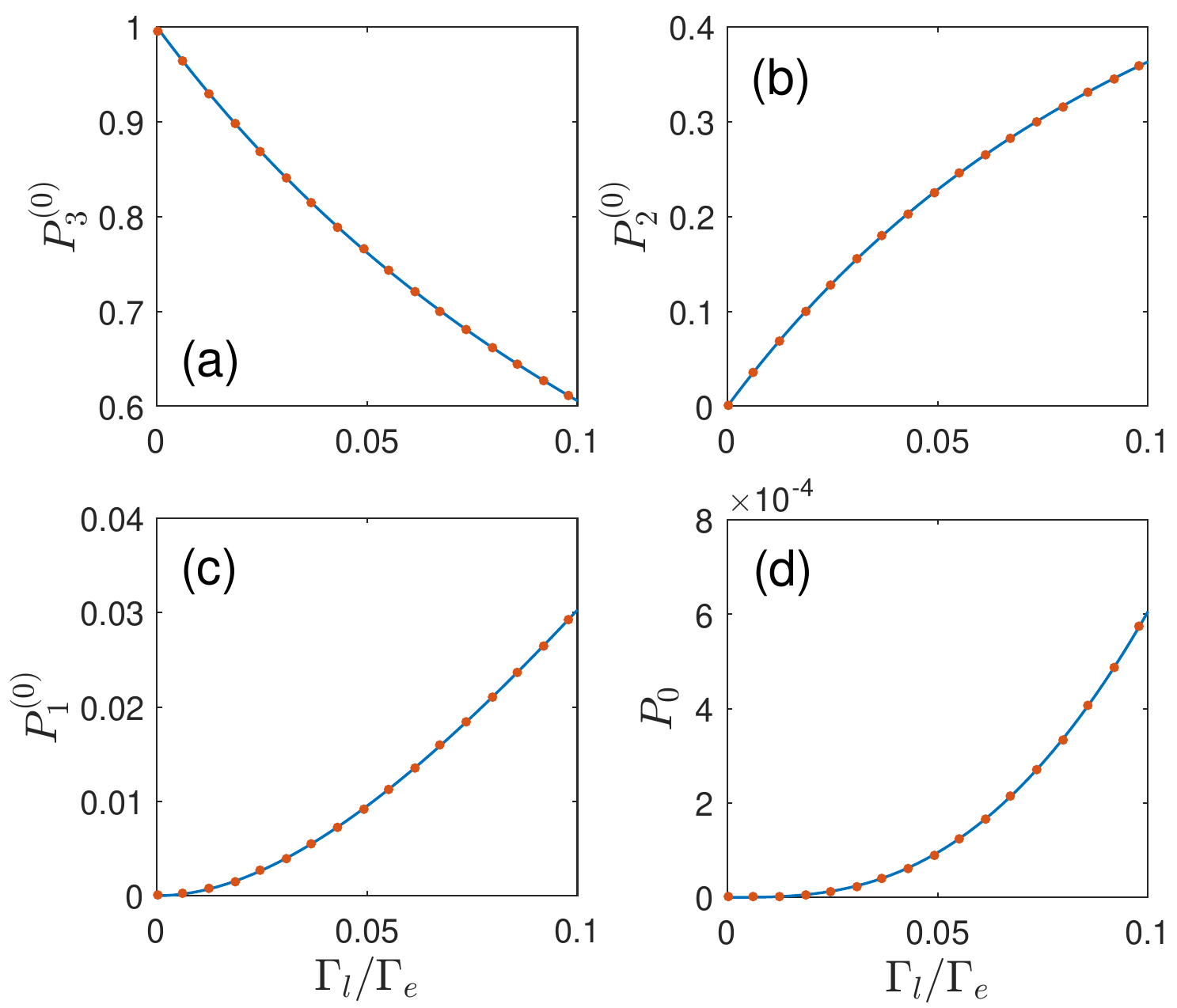}
\caption{(a)-(c) Total populations $P_N^{(0)}$ in the $N$th lowest-energy manifold of non-interacting states for $N = 1,2,3$ and (d) the vacuum population $P_0$ as a function of $\Gamma_l/\Gamma_e$, where $P_3^{(0)}$ corresponds to the target $N_T = 3$-particle Laughlin state population. The analytical results are shown by blue curves and the numerical ones by red dots. Pump and emission parameters are $\hbar\Gamma_p/\Delta = 5\times10^{-4}$ and $\hbar\Gamma_e/\Delta = 2.5\times10^{-4}$. Angular momentum potential strength is $\delta = 5\Delta$.
\label{Populations in the unperturbed manifold}}
\end{figure}

There are two main sources of fidelity loss in our scheme of Laughlin state preparation. The technically more challenging one to overcome is the already discussed loss to lower degenerate manifolds. We show in Fig. \ref{Populations in the unperturbed manifold} our numerical results for the variation of the total population in each of these manifolds for $N_T = 3$ as $\Gamma_l/\Gamma_e$ increases. The most detrimental contribution comes of course from the $N = 2$ states which brings about a decrease in the target Laughlin population in the first order of $\Gamma_l/\Gamma_e$. As the particle number increases this effect gets amplified as a result of increasing multiplicity of degenerate states in the lower manifolds, which is reflected in Fig. 3 of the main text. The remedy to reduce this imperfection is to achieve as low loss to emission ratio as possible.

\begin{figure}[htbp]
\includegraphics[scale=0.7]{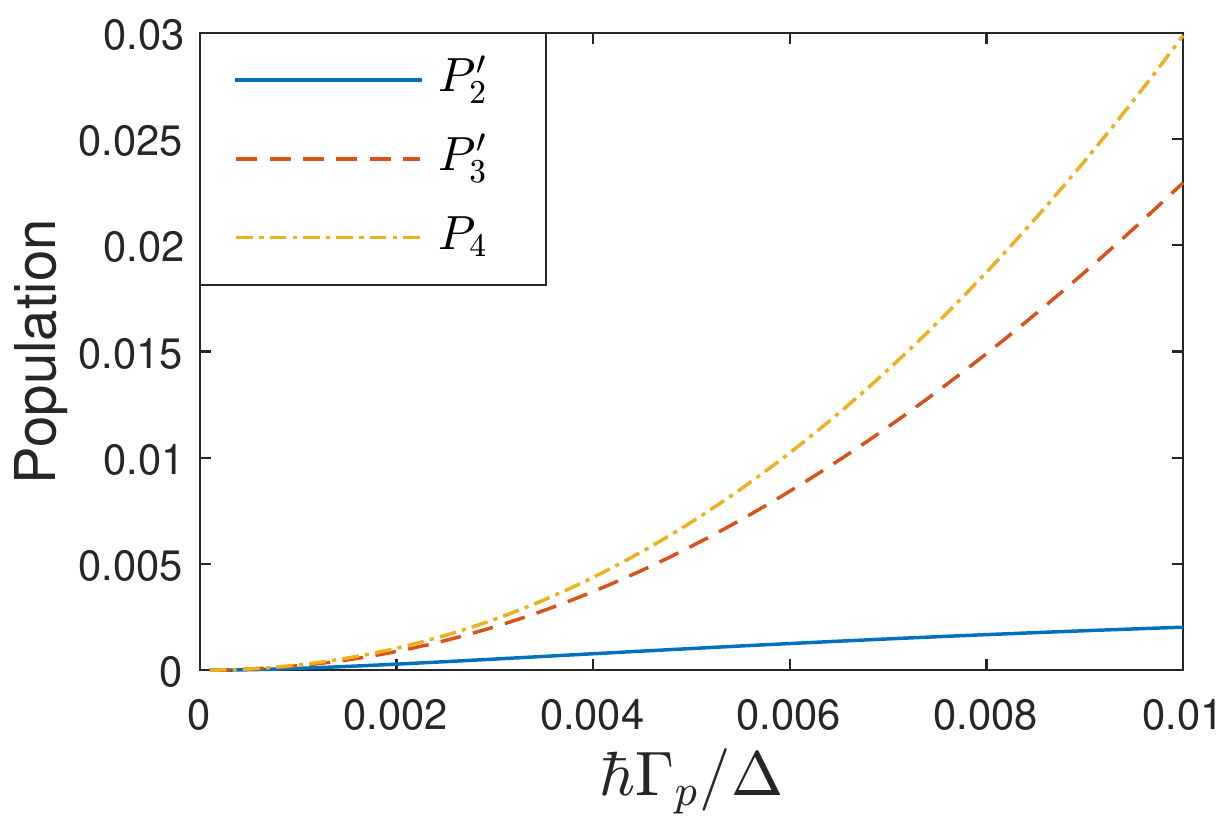}
\caption{Total populations $P_N^{\prime}$ outside the unperturbed lowest-energy manifold for $N = 2,3$ and the total population $P_4$ in the four-particle sector as a function of $\hbar\Gamma_p/\Delta$ with $\hbar\Gamma_e/\Delta = 5\times10^{-5}$ and $\Gamma_l/\Gamma_e = 2\times10^{-3}$, when the target state is the $N_T = 3$-particle Laughlin state. Angular momentum potential strength is $\delta = 5\Delta$.
\label{Populations outside the unperturbed manifold}}
\end{figure}

The other loss channel is the increase in the population of states lying outside of the degenerate manifolds as the pump linewidth $\Gamma_p$ increases. This effect is shown in Fig. \ref{Populations outside the unperturbed manifold} which is obtained by numerically solving the master equation for four particles when the target is the $N_T = 3$-particle Laughlin state. As can be seen from the figure, the four-particle ($N = N_T+1$) population dominates other non-degenerate state populations with $N<N_T$. This can be understood by looking at Eq. (\ref{Modified original relation}), which implies that when the difference $\Delta^{\prime} = \hbar(\omega_{N+1,N}-\omega_0)$ ($\omega_{at} = \omega_0$) is non-vanishing the emission rate will be suppressed. To make an order of magnitude estimate we assume that the populations are related through an approximate detailed balance condition $P_{N+1}\Gamma_l \propto P_N \Gamma_e/[1+(2\Delta^{\prime}/\hbar\Gamma_p)^2]\approx P_N \Gamma_e(\hbar\Gamma_p/2\Delta^{\prime})^2$. Since we expect most of the total population will be accumulated in the Laughlin state, the most populated states outside the degenerate manifolds would be the lowest-energy $(N_T+1)$-particle states. A tentative interpretation of these states may be in terms of a single quasi-particle excitation on top of a Laughlin state. Taking $\Delta^{\prime} \sim \Delta$ as in Fig. 2 of the main text and $P^{(0)}_3 \sim 1$ gives the correct order of magnitude for $P_4$ in Fig. \ref{Populations outside the unperturbed manifold}. 

\section{Details on the realization of the step-like potential in the angular-momentum basis}

In a recent work, it was shown that optical mode conversion may be realized by coupling a pair of multimode resonators together~\cite{stone2021optical}. The idea is to harness impedance matching: a lossless resonator transmits all light through it on resonance when the in-coupling through the input mirror is equal to the out-coupling through the output mirror; by employing a secondary \emph{cavity} as the output mirror, it is possible to achieve a situation where the transmission of the in-coupling mirror of the primary cavity is equal to the out-coupling of the TEM$_{00}$ of the primary cavity \emph{through} a higher-order mode of the second cavity, resulting in mode conversion.

\begin{figure}[h!]
	\centering
	\includegraphics[width=\columnwidth]{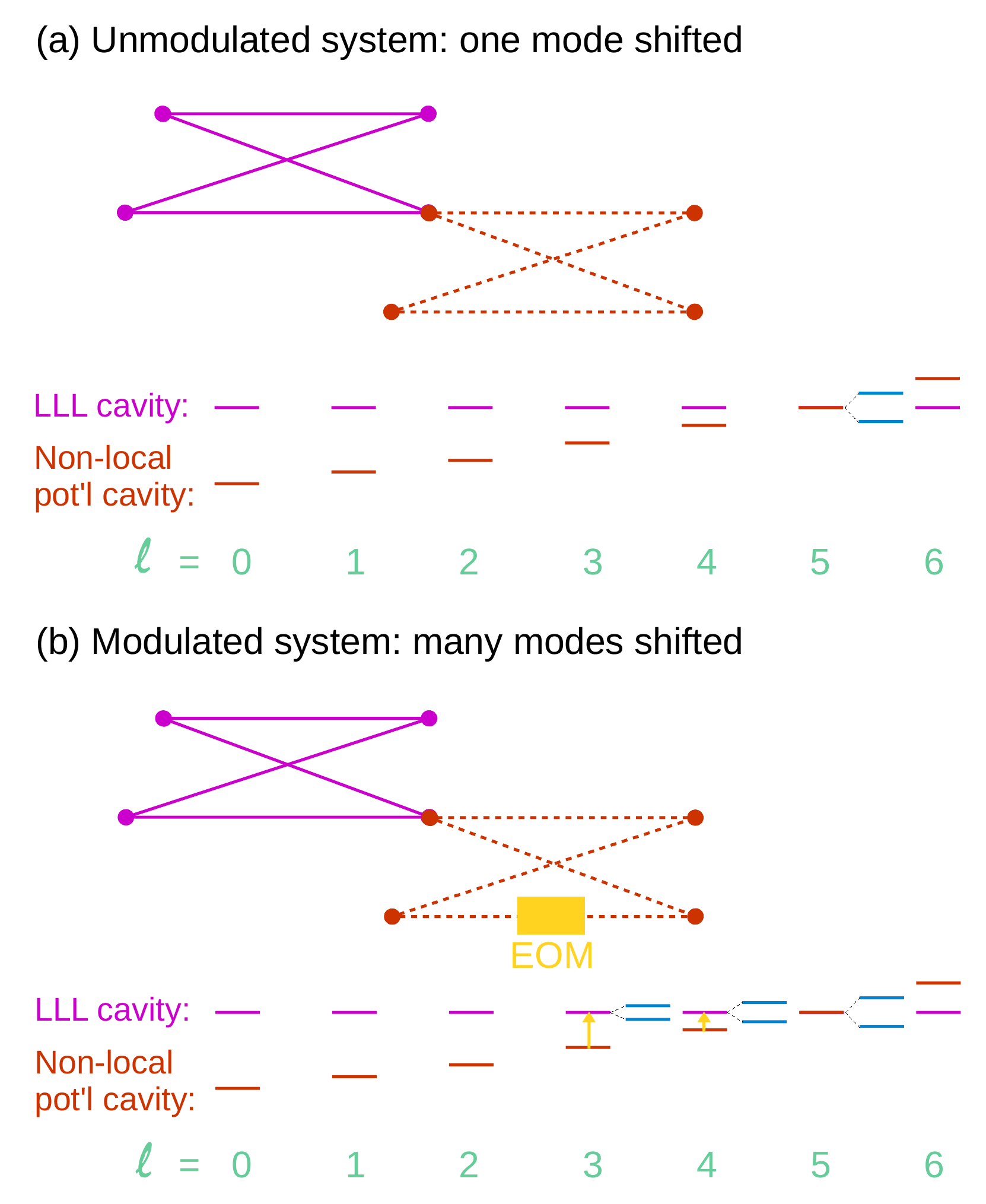}
	\caption{\textbf{Non-local potentials in coupled cavities.} We propose to employ a secondary resonator with an intracavity electro optic modulator (EOM) to control the frequencies of the modes of our LLL cavity. In the upper panel, we show a simplified approach, where the LLL cavity begins with all modes degenerate, and is coupled to a non-degenerate twisted cavity, one of whose modes is tuned to resonance with the LLL, there inducing hybridization and shifting the mode energy away. In the lower panel, we extend this idea to multiple modes by introducing the EOM into the secondary cavity, and modulating at appropriate frequencies to induce spectral weight of \emph{multiple} modes at the LLL energy, resulting in energy shifts for those modes.}
	\label{fig:Jon}
\end{figure}

What if the secondary cavity is single-ended, with no power leaking out of its other end (see Fig.~\ref{fig:Jon})? In this case, the secondary cavity does not induce extra losses but rather can be exploited to induce a mode-dependent phase-shift on the primary cavity, resulting in a mode-dependent energy shift. 

In the context of this work, the magnetic Hamiltonian in Eq.~(1) of the main text can be obtained using a primary cavity which is twisted to produce a flat lowest-Landau level as done in~\cite{schine2016synthetic,schine2019electromagnetic,clark2020observation}. 

As a secondary cavity, we propose to use a cavity that is twisted, but not to degeneracy. In this way, it is possible to length-tune a \emph{specific} LG- mode of the secondary cavity to resonance with the corresponding mode of primary cavity, thus shifting it away from the lowest Landau level. As it is shown in the upper panel of Fig.\ref{fig:Jon}, this provides an efficient way of tuning the frequency of a specific angular-momentum mode without affecting all other. 

As mentioned in the main text, our proposal requires blocking several angular momentum modes. Even though the resonance condition between the primary and secondary cavity modes only holds for a specific angular momentum value, the lower panel of Fig.~\ref{fig:Jon} shows how an electro-optical modulator (EOM) can be used to bridge the detuning. In this way, an effective resonance can be obtained for specific modes just by driving the EOM with the appropriate modulation frequencies, which enables to \emph{choose} which modes of the secondary cavity to tune away from resonance with the primary cavity.

The flexibility of this set-up allows to shape the step-like potentials in the angular momentum basis that are needed to confine fractional quantum Hall liquids and, if needed, introduce quasiholes in the liquid. Note that the efficiency of this method stems from the markedly non-local character of the potential in real space, which allows to overcome the limitations of the standard real-space potentials considered in previous works~\cite{Maca17,Umu17}.